\documentclass[letter]{aa} % for the letters 
\pdfoutput=1
\def\ga{\;\rlap{\lower 2.5pt\hbox{$\sim$}}\raise 1.5pt\hbox{$>$}\;}
\def\la{\;\rlap{\lower 2.5pt\hbox{$\sim$}}\raise 1.5pt\hbox{$<$}\;}
\usepackage{graphicx}
\usepackage{txfonts}
\begin{document}

   \title{ALMA detects a radial disk wind in DG Tau}

   %\subtitle{I. Overviewing the $\kappa$-mechanism}

   \author{M. G\"udel
          \inst{1},
          C. Eibensteiner\inst{1}, %\fnmsep\thanks{Just to show the usage of the elements in the author field}
          O. Dionatos\inst{1},
          M. Audard\inst{2},
          J. Forbrich\inst{3},
          S. Kraus\inst{4},
          Ch. Rab\inst{5},
          Ch. Schneider\inst{6},\\
          S. Skinner\inst{7},
          and E. Vorobyov\inst{1, 8}
          }

   \institute{University of Vienna, Dept. of Astrophysics,
              T\"urkenschanzstr. 17, 1180 Vienna, Austria,
              \email{manuel.guedel@univie.ac.at}
         \and
              Department of Astronomy, University of Geneva, Ch. d'Ecogia 16, 1290 Versoix, Switzerland
             %\email{other@e-mail1}
         \and
             University of Hertfordshire, Hatfield, Hertfordshire, United Kingdom
             %\email{other@e-mail2}
         \and
             University of Exeter, School of Physics and Astronomy, Stocker Road, Exeter, EX4 4QL, UK
             %\email{other@e-mail3}
         \and
             Kapteyn Astronomical Institute, University of Groningen, P.O. Box 800, 9700 AV Groningen, The Netherlands
             %\email{other@e-mail4}
         \and
             Hamburger Sternwarte, Gojenbergsweg 112, 21029 Hamburg, Germany
             %\email{other@e-mail2}
        \and             
             Center for Astrophysics and Space Astronomy, University of Colorado, 389 UCB, Boulder, Colorado 80309-0389, USA
             %\email{other@e-mail5}
             %\thanks{The university of heaven temporarily does not accept e-mails}
        \and
             Research Institute of Physics, Southern Federal University, Stachki Ave. 194, Rostov-on-Don, Russia
             }
\authorrunning{M. G\"udel et al.}

   \date{Received September 18, 2018; accepted October 20, 2018}

% \abstract{}{}{}{}{} 
% 5 {} token are mandatory
 
  \abstract
   {}
   {
   We aim to use the
high spatial resolution of the Atacama Large Millimeter/submillimeter Array (ALMA) to map the
   flow pattern of molecular gas near DG Tau and its disk, a young stellar object driving a jet and a molecular 
   outflow.
  }
   {We use observations from ALMA 
   in the $J=2-1$ transition of $^{12}$CO, $^{13}$CO, and C$^{18}$O to study 
   the Keplerian disk of DG Tau and outflows that may be related to the disk and the jet. 
   }
   {We find a new wind component flowing radially at a steep angle ($\approx 25^{\circ}$ from the vertical) above 
    the disk with a velocity of $\approx 3.1$~km~s$^{-1}$. 
    It continues the trend of decreasing  velocity for increasing distance from the jet axis (``onion-like velocity structure'').}
   {The new component is located close to the protostellar disk surface and may be related to photoevaporative winds. }


   \maketitle
%
%-------------------------------------------------------------------

\section{Introduction}

Protostellar outflows, jets, and disk winds play important roles 
in disk accretion, dispersal, and angular momentum transport 
\citep{frank14}. Observations indicate a physical link between
accretion and ``ejection'', the mass outflow from a disk typically amounting
to a few percent of the accretion rate onto the star \citep{white04}. 
While collimated atomic jets reach velocities up to several 100~km~s$^{-1}$,
molecular outflows are slower (up to a few
tens of km~s$^{-1}$) and less well collimated. They may result from entrainment by 
the fast jet interacting with the disk material and the surrounding protostellar envelope
\citep{dionatos17}. Slow disk winds may also result  
from magnetocentrifugal acceleration
(with velocities of a few times the sound velocity at a few scale heights above the disk; 
\citealt{bai13}) and/or
from X-ray/ultraviolet disk irradiation 
from the central star \citep{ercolano08, gorti09}. Such photoevaporative winds are typically 
launched at a few tens of astronomical units from the star. 
\citet{kitamura96a} (hereafter K96) proposed a  process wherein a strong stellar wind interacts with the disk surface to drive a 
near-radial expansion and therefore erosion of the disk. In this  Letter we present ALMA
 observations that trace disk mass loss in  the young stellar object DG Tau.

\section{The target: DG Tau}

DG Tau  
is, judging from its ``flat'' spectrum, at a transition between a Class-I protostar and a classical T Tauri star
\citep{pyo03, calvet94}. It is surrounded by a disk of  gas and dust \citep{dutrey96, kitamura96a, kitamura96b, testi02}, 
with a size in the millimeter continuum of about $1\farcs 1\times 0\farcs 6$ (133~au $\times$ 73~au;
\citealt{dutrey96}). DG Tau ejects a well studied collimated jet 
with velocities of several 100~km~s$^{-1}$ (e.g., \citealt{lavalley97, eisloeffel98, white14}). 
We use a distance to DG Tau of 121.2 (119.1-123.4)~pc measured by GAIA
\citep{gaia2016, gaia2018}, a value much smaller than hitherto assumed ($d \approx 140$~pc).

\citet{kitamura96a}  mapped DG Tau's large-scale environment in $^{13}$CO (1--0), revealing molecular 
gas in an extended disk-like structure with a radius of $\sim 2800$~au (for $d = 140$~pc) showing, 
instead of a rotational velocity pattern, radial expansion with a velocity of $\sim 1.5$~km~s$^{-1}$. 
\citet{testi02} additionally mapped the Keplerian motion of the inner disk. 
Higher-excitation $^{12}$CO (6--5) and $^{12}$CO (3--2) observations presented earlier by \citet{schuster93}
indicated a narrow line, cut off by blueshifted self absorption, plus extended wings (of a few km~s$^{-1}$) indicating 
an outflow. 

\citet{takami04} and \citet{beck08} identified a cone-like structure of outflowing gas surrounding the jet axis  
in near-infrared H$_2$ emission, with a radius at its distant end of about 40~au and a flow velocity of about 15~km~s$^{-1}$. 
\citet{agraamboage14} suggested a hollow-cavity geometry with a half-opening angle of 30$^{\circ}$
and velocities down to a few kilometers per second at larger distances from the axis; the structure was furthermore observed in ultraviolet 
H$_2$ fluorescence lines by \citet{schneider13} using the Hubble Space 
Telescope. Overall, a picture is emerging in which the jet/outflow/wind system shows 
an onion-like structure around the axis with decreasing velocity for increasing distance from the axis 
\citep{takami04, agraamboage14}.

\section{Observations}

Our observations of DG Tau  were obtained by the Atacama Large 
Millimeter/submillimeter Array (ALMA) in Cycle 3 on 24/26 September 2016 using 42 of the 
12m antennas, providing an angular resolution (beam size) of $\sim$0$\farcs 21\times 0\farcs 12$ 
(25~au $\times$ 15~au). The observations in band 6 comprise the transitions of 
$^{12}$CO $J = 2-1$ (230.538~GHz),
$^{13}$CO $J = 2-1$ (220.399~GHz), and
C$^{18}$O $J = 2-1$ (219.560~GHz). We also included the SiO $J = 5-4$ (217.105~GHz) 
transition in the setup but did not obtain any detection. The spectral bandwidth was 117 MHz (corresponding to 
a velocity range of $\sim$152--162 km~s$^{-1}$ per sideband), and the spectral resolution 
was 0.122 MHz (velocity resolution of $\sim$0.159--0.167 km~s$^{-1}$).
The on-source time amounted to 1.95~h. Three extragalactic calibrators were used (quasars J0238+1636, 
J0510+1800, J0403+2600).
We applied standard procedures provided by the CASA software version 4.7.0-1 for data reduction, 
calibration, and data extraction. The line data presented here were continuum-subtracted and
cleaned 
following prescriptions in the CASA User Reference \&  Cookbook, public release 2017. Typical background 
noise levels were 3.8, 4.3, and 3.1~mJy/beam 
for $^{12}$CO, $^{13}$CO, and C$^{18}$O, respectively. Velocities are given relative to the local standard
of rest (lsr) unless  noted otherwise (Sect.~\ref{discussion})

   \begin{figure}[t!]
   \centering
   \vskip -0.3truecm\vbox{
   \includegraphics[angle=-90,width=8.6cm,  trim={0 0 0 0}, clip]{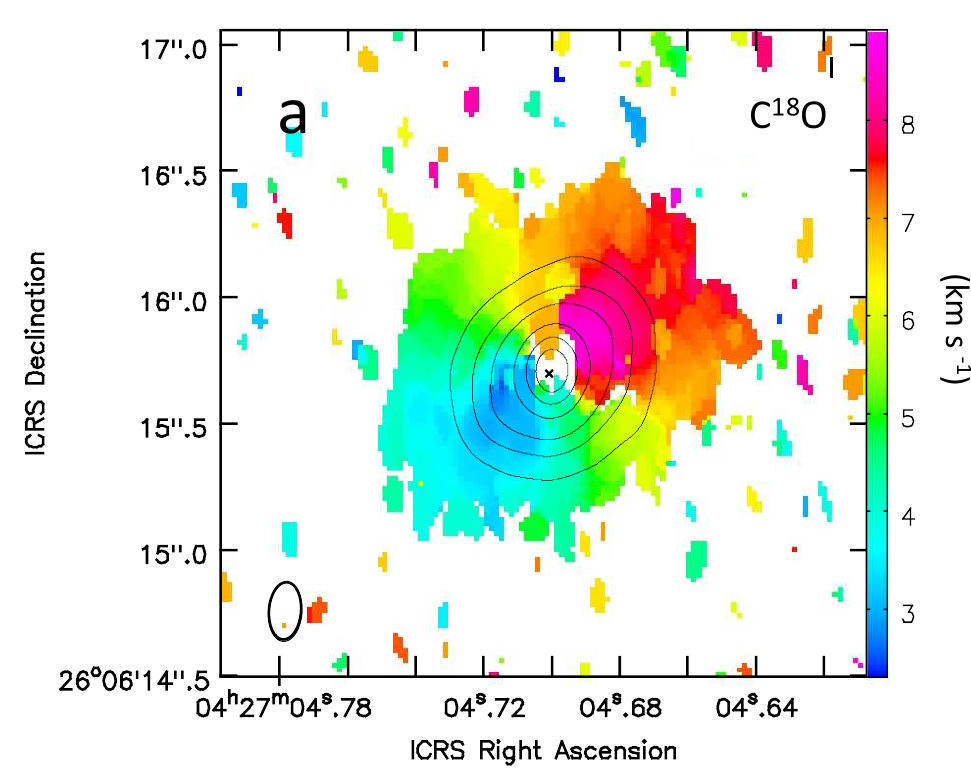}
   \includegraphics[width=8.8cm,  trim={0 0 0 0}, clip]{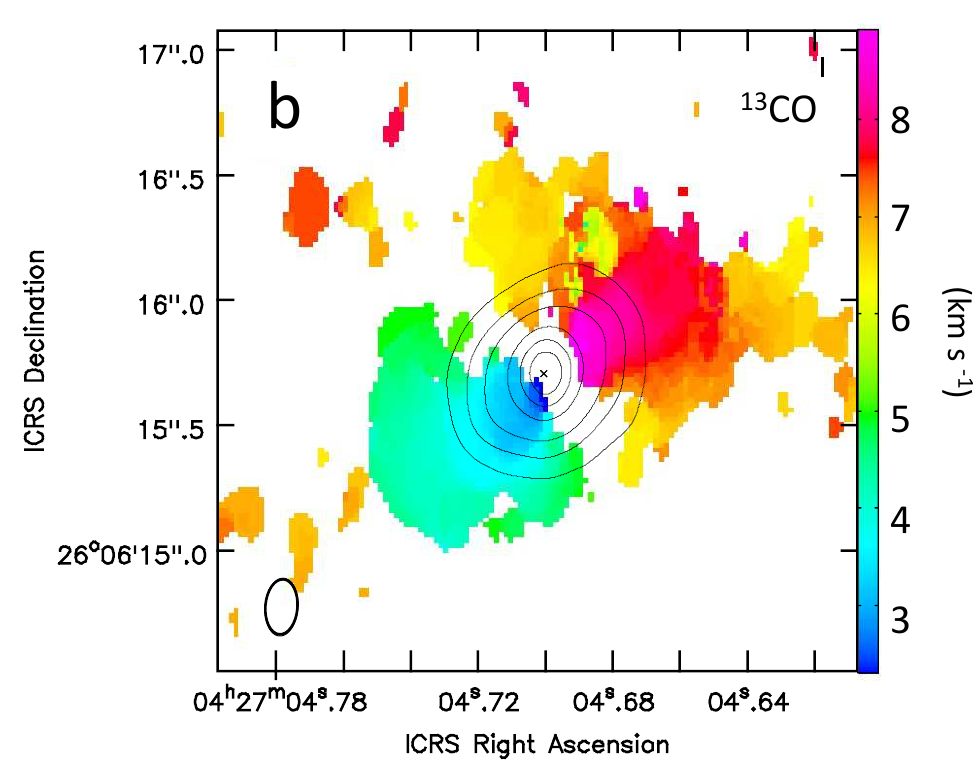}
   \hbox{\hskip 0.3truecm\includegraphics[width=8.0cm,  trim={0 0 0 0}, clip]{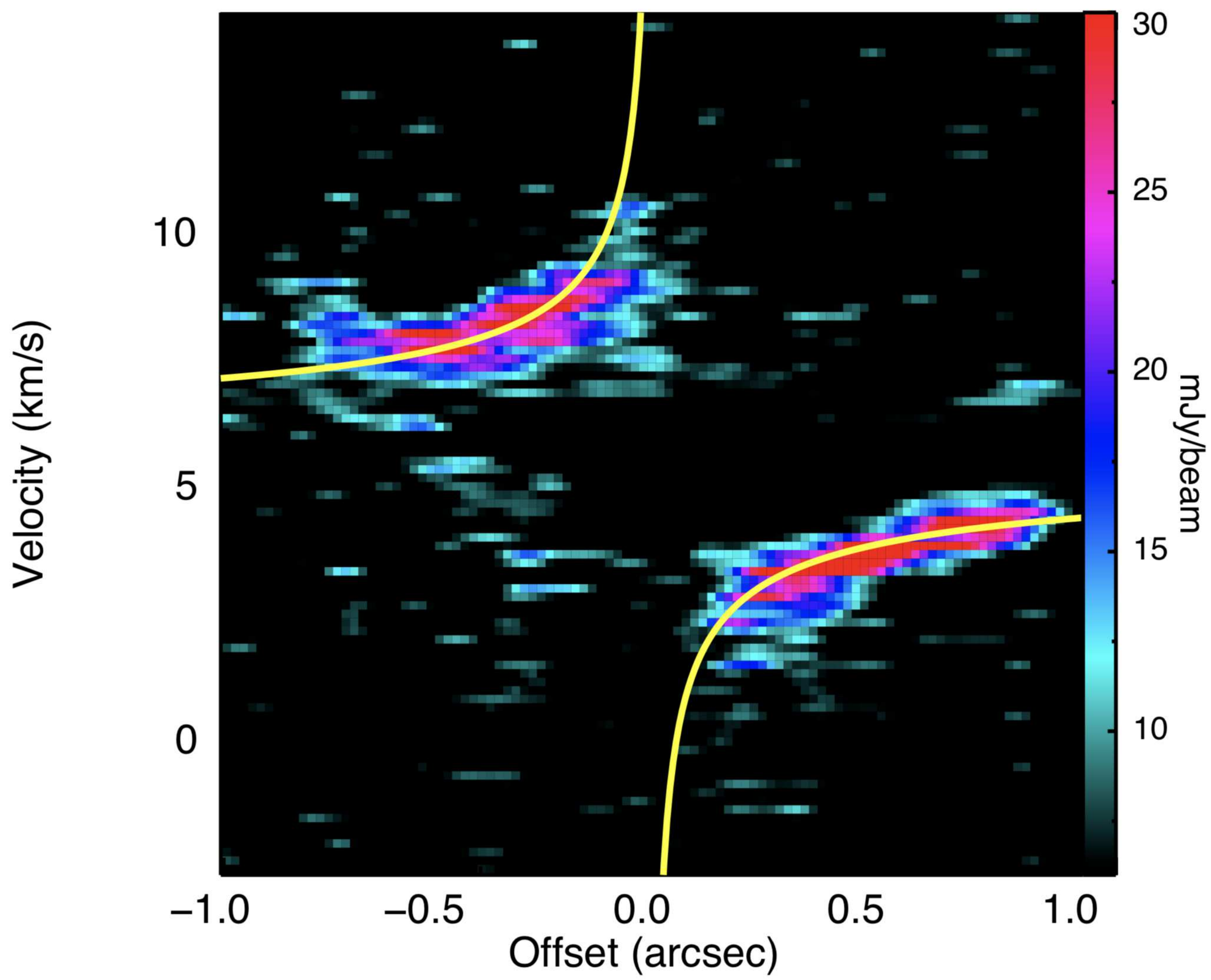}}
   }
      \caption{\textit{Top and center}: First-moment maps for C$^{18}$O (top; $v_{\rm lsr}$=2.3--8.9~km~s$^{-1}$) 
      and $^{13}$CO (middle; $v_{\rm lsr}$=1.5--8.2~km~s$^{-1}$). Beam size is  
      $0\farcs 23$$\times$$0\farcs 13$ (ellipses at lower left). 
      Contours mark the continuum emission levels of 9.4, 15.1, 20.8, 26.5, 37.8, 49.2~mJy/beam, with a 
      peak at 60.6~mJy/beam. The cross symbol marks the position of the star. 
      \textit{Bottom}: Position-velocity diagram  
      along  major axis of projected 
      disk in $^{13}$CO; position angle 134.3$^{\circ}$. Yellow curves represent the Keplerian profiles for a 
      0.5$M_{\odot}$ star.
              }
         \label{fig:firstmoment}
   \end{figure}

\section{Results}\label{results}

Figure~\ref{fig:firstmoment} shows first-moment (flux-weighted velocity) maps with the color coding given 
in the bar on the right.   The gas emission extends to larger radii than the dust continuum disk (overplotted 
in contours) as is typically observed (e.g., \citealt{ansdell18, facchini17}).
The C$^{18}$O (a) and $^{13}$CO (b) maps indicate a 
symmetric disk compatible with an inclination of $\sim 38^{\circ}$ 
\citep{eisloeffel98}.  The position-velocity diagram 
(Fig.~\ref{fig:firstmoment}c) extracted along the disk major axis of the  $^{13}$CO map  
agrees with a Kepler rotation profile for a star of 0.5 solar masses ($M_{\odot}$; solid lines; similar
for $^{18}$CO).
These figures indicate a systemic velocity relative to lsr of
$v_{\rm sys}\approx 5.5$~km~s$^{-1}$, in agreement with \citet{testi02} and K96.
 Judging from the jet geometry  (the approaching jet points toward a position angle
of $\sim$222~deg, that is, to the southwest (SW), \citealt{lavalley97}), the southeast (SE) half of the disk is 
approaching, the disk rotates clockwise, and thus the northeast (NE) half-disk is in front.
Around zero velocity relative to $v_{\rm sys}$ (NE-SW diagonal) the $^{13}$CO disk emission is 
subject to absorption, supporting reports by \citet{schuster93} (line profile) and \citet{testi02} (channel maps).

   \begin{figure*}
   \centering
   \vbox{
   \includegraphics[width=18cm, trim={50 15 90 85}, clip]{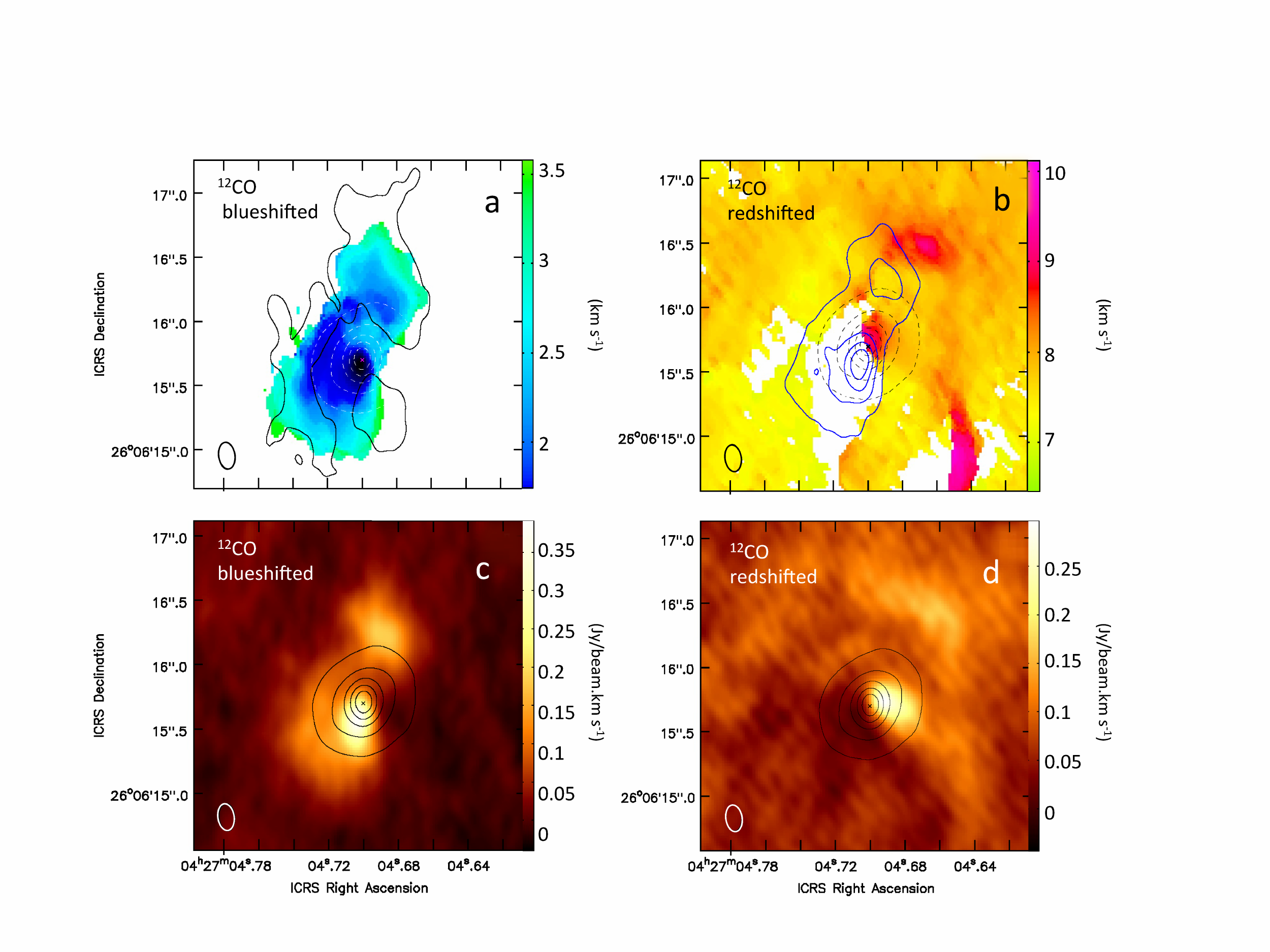}}
   \vskip -0.3truecm
      \caption{{\bf Top row:} $^{12}$CO first-moment maps - {\bf (a)} \textit{Left}: 
      Blueshifted, $v_{\rm lsr} = $ -1.2 -- +3.6~km~s$^{-1}$ (color; black below 1.6~km~s$^{-1}$) 
      and 4.5 -- 4.8~km~s$^{-1}$ (so\-lid outline). Dashed contours represent the continuum emission 
      (4.3, 15.1, 26.5, 37.8, 49.2~mJy/beam, peak 60.6~mJy/beam). The beam size is $0\farcs 21$~$\times$~$0\farcs 12$. 
      {\bf (b)} \textit{Right}:  Redshifted,  $v_{\rm lsr} = 7.0-10.1$~km~s$^{-1}$. Solid contours represent
      blueshifted intensity from (c). Dashed contours and beam size are as in (a).
      {\bf Bottom row:} Intensity maps - {\bf (c)} \textit{Left}: Blueshifted intensity 
      ($v_{\rm lsr} =$ -1.2 -- +4.8~km~s$^{-1}$). Beam size is $0\farcs 21\times 0\farcs 12$.
      Background rms $\approx 9.6$~mJy/beam~km~s$^{-1}$, peak flux = 0.39~Jy/beam~km~s$^{-1}$.  
      {\bf (d)} \textit{Right}: Redshifted intensity 
       ($v_{\rm lsr} =$ +7.0 -- +11.4~km~s$^{-1}$). Background rms $\approx 12$~mJy/beam~km~s$^{-1}$, 
       peak flux  = 0.28~Jy/beam~km~s$^{-1}$. Beam size is as in (c), 
       solid contours in (c) and (d) show the same as the dashed contours in (a) and (b).}\label{fig:12CO}
   \end{figure*}

The $^{12}$CO first-moment map, in contrast, reveals complex velocity structure presented 
separately for the blue- and  redshifted channels in Fig.~\ref{fig:12CO}. 
The blueshifted emission (a) is concentrated toward SE-E-NE-N-NW of the center for $v < 3.5$~km~s$^{-1}$, 
while redshifted emission (b) is seen  almost everywhere for $v > 7$~km~s$^{-1}$. The 3.5--5.5~km~s$^{-1}$ 
range is almost completely absorbed probably
by cool foreground material similar to the $^{13}$CO map (Fig.~\ref{fig:firstmoment}b) except for the 
4.5--4.8~km~s$^{-1}$ interval. The key features of the $^{12}$CO maps are  (labeled in Fig.~\ref{zoom} and
Fig.~\ref{redlarge}) as follows.
\begin{enumerate}
\item Within the projection of the continuum disk, the Keplerian disk is evident, 
      especially near the disk center (strongly redshifted in the NW 
      and blueshifted in the SE).
\item The Keplerian disk emission is asymmetric as bright redshifted and blueshifted intensity is
      missing toward  N and E, respectively. 
\item Additional blueshifted emission appears toward the E and N (Fig~\ref{fig:12CO}a), the 
      highest velocities  apart from those in the disk center (dark blue) revealing an arc-like structure from SE 
      to NW, approximately along the outermost two contours of the continuum disk (apart from additional blueshifted emission). 
      The lowest-velocity blueshifted emission is seen in the 4.5--4.8~km~s$^{-1}$ window (solid contours in 
      Fig.~\ref{fig:12CO}b) and reaches farther away from the star, also to SW--NW.
\item Much of the region outside the disk area shows redshifted emission, with
      two high-speed regions toward NW and SW (red,  feature [5] in Fig.~\ref{redlarge}).
\end{enumerate}

\section{Discussion}\label{discussion}

Feature [1] corresponds to the appearance of the first-moment map for $^{13}$CO 
(Fig.~\ref{fig:firstmoment}b) where absorption suppresses emission from two wedge-shaped regions along the NE-SW diagonal, 
also seen for low velocities in $^{12}$CO along the minor axis (see also Fig.~\ref{fig:12CO}c,d). Although parts of the outer disk can be traced out to $0\farcs 5$-1$^{\prime\prime}$, the Keplerian 
disk is very centrally brightened in a region with high radial velocities. However, in Figs.~\ref{fig:12CO}c,d 
both the redshifted and blueshifted high-velocity inner Keplerian disk emission regions are asymmetric [2] with respect to 
the major axis of the projected disk. The emission is almost completely absent in the NE half of the disk (which 
is closer to the observer). This feature is obviously related to the line of sight. We suggest that the specific viewing
angles result in additional absorption; we emphasize that a wide range of (Keplerian) velocities 
($\pm$ several km~s$^{-1}$) within a small area around the disk center are affected. We speculate that disk flaring leads to much longer 
line-of-sight paths through the local upper atmosphere on the closer half of the disk, perhaps leading to local 
``self-absorption'' of the emission without velocity difference. The line-of-sight angle is  less inclined with regard
to the disk surface normal on the opposite 
side of the disk, meaning that the emission suffers less absorption.  The flaring index of the relevant layer (CO and/or dust) 
is not known, making a quantitative estimate challenging, but previously used indices of 1.2 and 1.25 (\citealt{guilloteau11} and 
\citealt{podio13}, respectively) give similar results and suggest a height of 10~au above the mid-plane at $r =$ 69~au,
roughly compatible with our estimate for the height of the wind structure derived below. Disk flaring 
could thus be a significant factor for absorption. 

Feature [3] is the most important finding of these 
observations, revealing a mass flow that is not compatible with disk rotation but  
appears to flow radially outward above the disk surface. This material must be located 
in front of the disk.

\begin{figure}[h!]
\hskip -1.5truecm\includegraphics[width=10.2cm, trim={0 0 0 70}, clip]{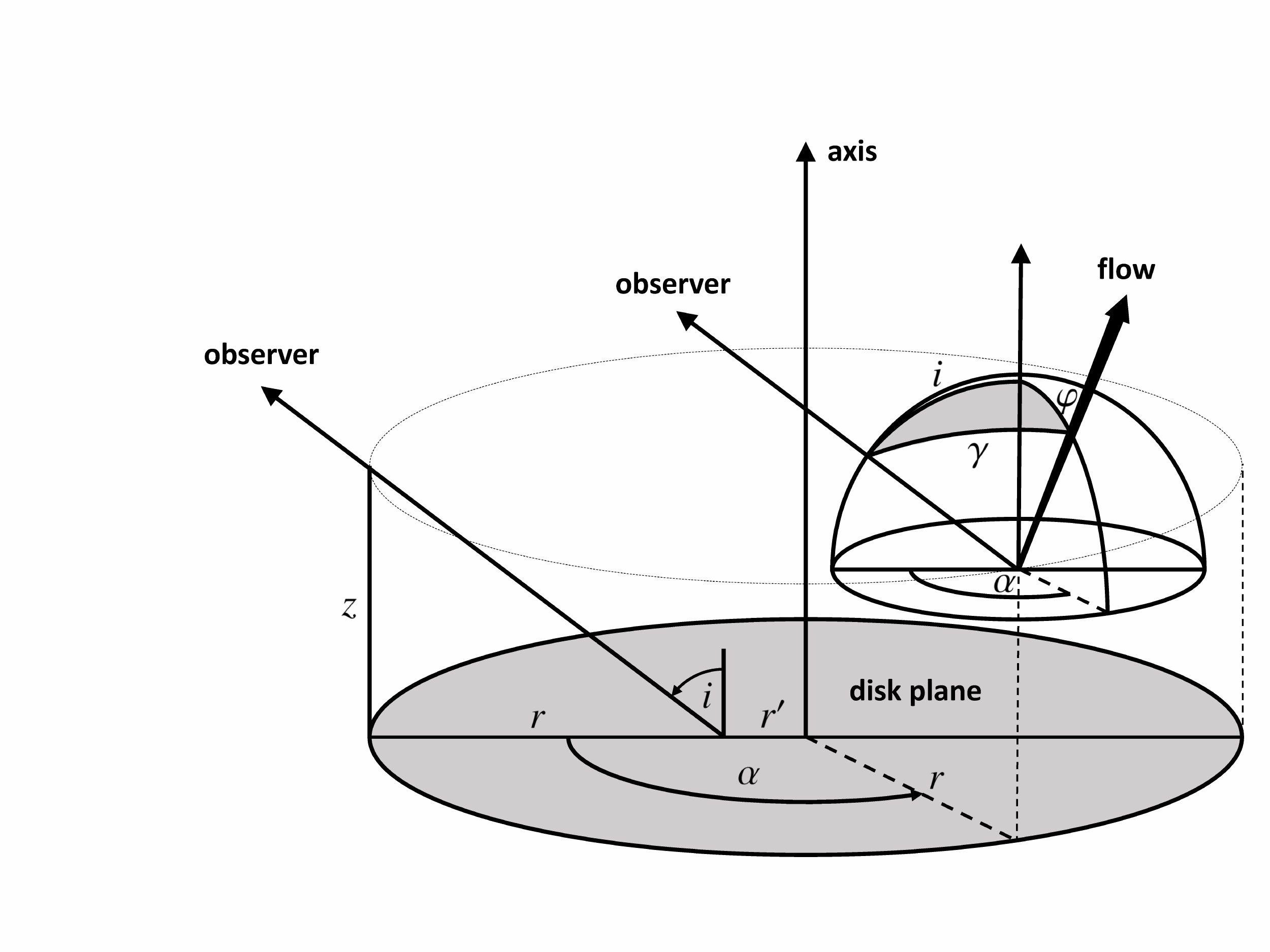}
\vskip -0.5truecm
\caption{Geometry of the flow relative to the disk and the line of sight (see main text and Appendix for details).}
\label{geometry_main}
\end{figure}

We estimate the height of the emitting structure and its flow direction using Fig.~\ref{geometry_main}  (see also 
Appendix~\ref{sect:geometry}), adopting a disk inclination of
$i = 38^{\circ}$ against the line of sight \citep{eisloeffel98}. We henceforth use velocity offsets relative
to the systemic velocity $v_{\rm sys} = 5.5$~km~s$^{-1}$ and define an azimuth angle $\alpha$ measured on the
disk clockwise from the NE minor axis. In Fig.~\ref{fig:12CO}a,
the blueshifted arc is nearly concentric to the continuum contours. The distance of the fast 
flow ($v \la$--3~km~s$^{-1}$) from the disk cen\-ter along the major axis is $\approx$$0\farcs 45\pm 0\farcs 12$. 
If the arc structure is located at a height $z$ above the disk plane, then the line of sight through the arc at 
$\alpha = 0^{\circ}$ cuts the disk plane on the minor axis at a distance of $r^{\prime} = r - z\tan i$ 
from the disk center. Projection
owing to the disk inclination compresses    $r^{\prime}$  to $r^{\prime}_{\rm proj} =  r^{\prime}\cos i = r\cos i - z\sin i$.
We estimate $r^{\prime}_{\rm proj} = 0\farcs 23 - 0\farcs 44$. 
From 
\begin{equation}\label{projection}
{z\over r} = {\cos i - r^{\prime}_{\rm proj}/r\over \sin i}
\end{equation}
and the requirement $z\ge 0$, we find $z/r = 0.03-0.15$, that is, the flow emission is located  immediately above
the disk midplane ($z \la 10$~au for $r = 69$~au [$0\farcs 57$]).

The flow is detected only in the NE half of the disk (around $\alpha \approx 0^{\circ} \pm 90^{\circ}$), indicating that 
it is inclined outwards in such a way that the radial velocity at larger $\alpha$ 
is small enough for the flow to be subject to foreground absorption. We assume a flow velocity of $v$ at an angle 
of $\varphi$ to the local vertical  on the disk midplane (Fig.~\ref{geometry_main}), away from the axis. At an azimuth $\alpha$, the spherical triangle 
between  the local vertical, the line of sight, and the flow direction (at angle $\varphi$ to the vertical in
azimuth direction $\alpha$) requires that for the  angle  $\gamma$ between the flow and the line of sight,
\begin{equation}\label{proj}
\cos \gamma = \cos i\cos\varphi + \sin i\sin\varphi\cos\alpha
,\end{equation}
and the radial flow velocity is  $v_{\rm rad} = v\cos\gamma$.
From Fig.~\ref{fig:12CO}a, at the nearest point on the minor axis, with azimuth $\alpha_1 = 0$, 
the radial velocity is  $v_{\rm rad}(\alpha_1) \approx -3$~km~s$^{-1}$. We estimate that for 
$\alpha_2 = 105^{\circ}$ (toward NW), the radial flow velocity reaches  $v_{\rm rad}(\alpha_2) = -2$~km~s$^{-1}$ 
(and is absorbed at larger $\alpha$ and presumably smaller velocity offsets);  
the ratio $v_{\rm rad}(\alpha_2)/v_{\rm rad}(\alpha_1) = 2/3$ determines, through Eq.~\ref{proj} for the numerator 
and denominator, the value of $\varphi$. We obtain $\varphi \approx 25^{\circ}$. The flow close to the disk is therefore 
relatively steep; this geometry ($\varphi < i$) explains why the fast blueshifted flow does not expand  much to the NE. 
The full velocity is about $v = v_{\rm rad}(\alpha_1)/\cos(\varphi-i) \approx  -3/\cos(-13^{\circ}) = -3.1$~km~s$^{-1}$. 
At an azimuth of $\alpha_3 = 180^{\circ}$,  we expect a radial flow velocity of $v_{\rm rad}(\alpha_3) =  v\cos(\varphi+i) 
\approx -1.4$~km~s$^{-1}$ relative to $v_{\rm sys}$, that is, 4.1~km~s$^{-1}$ relative to 
lsr, which lies in the middle 
of the fully absorbed spectral region.

One possibility for the physical origin of this wind is photoevaporation driven by ultraviolet, extreme ultraviolet, 
or X-ray irradiation from the central star. Both the velocity and location of the feature agree with predictions for 
photoevaporative winds while all known magnetocentrifugal wind features known in DG Tau are at significantly higher 
velocities and are much closer to the star. Predictions for observable features of photoevaporative winds concentrated 
on atomic or ionized species and it is difficult to estimate if heating is sufficient to lift the deeper disk layers 
where CO molecules abound. We used the resulting temperature structure from a thermochemical 
disk model of DG Tau published by \citet{podio13} that considers the high-energy radiation of DG Tau. We then determined 
the layer at which the escape temperature is reached using DG Tau's high-energy luminosity (where 
$T_{\rm esc} = Gm_{\rm H}M_{*}/[kr]$, $G$ = gravitational constant, $m_{\rm H}$ = mass of 
hydrogen atom, $M_{*}$ = stellar mass, $k$ = Boltzmann constant, $r$ = distance from star; \citealt{ercolano08}). 
This rough model predicts escape at $r=60-70$~au, significantly above the molecular layer, but the exact model strongly
depends on details such as the spectrum of the irradiation, the disk structure (e.g., flaring), and
disk density. This requires a dedicated thermochemical model considering the new spatially resolved data 
(i.e., C$^{18}$O for the disk structure).

The widely distributed redshifted emission [4] was already noted by \citet{testi02} and is probably related to the 
molecular gas in the larger region around DG Tau. We note that the redshifted $^{12}$CO line remains, expectedly, unaffected 
by  the blueshifted flow in front, given the frequency shift of the line.
Specifically, we note a strongly redshifted arc-like feature  ([5] in Fig.~\ref{redlarge}) approximately 
1$^{\prime\prime}$ SW of the disk with a maximum  radial velocity of about +4.5~km~s$^{-1}$ (relative to $v_{\rm sys}$; 
Fig.~\ref{fig:12CO}b). This feature continues down to the central disk region at decreasing radial velocity 
(Fig.~\ref{fig:12CO}d). We speculate that this is an accretion 
flow from the remnant envelope in front of the disk  spiraling toward the disk, similar to streams detected by ALMA in 
the protostar L1489 IRS \citep{yen14};  the variation in redshift would
then be due to projection.  Another detached redshifted feature is seen to the N.

\section{Conclusions}
Our ALMA observations uncover a new outflow component in DG Tau.
Our geometric estimates show that the wind may reach a velocity of about 3.1~km~s$^{-1}$, to be compared with 
$\sim 1.5$~km~s$^{-1}$ at the larger distances reported by K96, supporting the picture of an 
``onion-like'' structure of flows with a velocity gradient increasing toward the axis. 
The wind emission originates close to the disk surface at a disk radius of approximately 40--70~au and is 
inclined outward by about 25$^{\circ}$ from the vertical. The size of this structure is much larger than
the previously described H$_2$ cone (e.g., \citealt{takami04, schneider13}).  We speculate that we 
see CO gas entrained by a photoevaporative flow \citep{ercolano08, gorti09} driven outward either by inclined magnetic fields 
(for weakly ionized gas) or the action of the stellar wind. For large distances (50--100~au) in the early disk stages, 
far-ultraviolet-driven
photoevaporation is particularly attractive \citep{gorti09}  although model calculations have so far favored flows of
atomic gas. Whether or not the flow we are observing here is related to the 
radial flow reported by K96 for much larger distances remains unclear.

\begin{acknowledgements}
SK acknowledges support from ERC grant No.\ 639889. EV acknowledges support from the Russian 
Ministry of Education and Science project 3.5602.2017.
This paper makes use of the following ALMA data: ADS/JAO.ALMA\#2015.1.00722.S. ALMA 
is a partnership of ESO (representing its member states), NSF (USA) and NINS (Japan), 
together with NRC (Canada), NSC and ASIAA (Taiwan), and KASI (Republic of Korea), in 
cooperation with the Republic of Chile. The Joint ALMA Observatory is operated by 
ESO, AUI/NRAO and NAOJ.
This work has made use of data from the European Space Agency (ESA) mission
{\it Gaia} (\url{https://www.cosmos.esa.int/gaia}), processed by the {\it Gaia}
Data Processing and Analysis Consortium (DPAC,
\url{https://www.cosmos.esa.int/web/gaia/dpac/consortium}). Funding for the DPAC
has been provided by national institutions, in particular the institutions
participating in the {\it Gaia} Multilateral Agreement.
\end{acknowledgements}


\begin{thebibliography}{}

 \bibitem[Agra-Amboage et al.(2014)]{agraamboage14}Agra-Amboage, V., Cabrit, S., Dougados, C., et al. % Kristensen, L.~E., Ibugi, L., Reunanen, J. 
          2014, A\&A, 564, A11
 \bibitem[Ansdell et al.(2018)]{ansdell18}Ansdell, M., Williams, J.~P., Trapman, L., et al. 2018, ApJ, 859, id.~21
 \bibitem[Bai \& Stone(2013)]{bai13}Bai, X.-N., Stone, J.~M. 2013, ApJ, 769, 76
 \bibitem[Beck et al.(2008)]{beck08}Beck T.~L., McGregor, P.~J., Takami, M., Pyo, T.-S.  2008, ApJ, 676, 472
 \bibitem[Boehler et al.(2017)]{boehler17}
          Boehler, Y., Weaver, E., Isella, A., et al. 2017, ApJ, 840, id. 60
 \bibitem[Calvet et al.(1994)]{calvet94}Calvet, N., Hartmann, L., Kenyon, S.~J., Whitney, B.~A. 1994, ApJ, 434, 330
 \bibitem[Cleeves et al.(2016)]{cleeves16} Cleeves, L.~.I., \"Oberg, K.~I., Wilner, D.~J., et al. 2014, ApJ, 832, 110 
 \bibitem[Dionatos \& G\"udel(2017)]{dionatos17}Dionatos, O., G\"udel, M. 2017, A\&A, 579, A64
 \bibitem[Dougados et al.(2000)]{dougados00}Dougados, C., Cabrit, S., Lavalley, C., M\'enard, F. 2000, A\&A, 357, L61
 \bibitem[Dunham et al.(2014)]{dunham14}Dunham, M.~M,, Vorobyov, E.~I., Arce, H.~G. 2014, MNRAS, 444, 887     
 \bibitem[Dutrey et al.(1996)]{dutrey96}Dutrey, A., Guilloteau, S., Duvert, G., et al. 1996, A\&A, 309, 493 
 \bibitem[Eisl\"offel \& Mundt(1998)]{eisloeffel98}Eisl\"offel, J., Mundt, R. 1998, AJ, 115, 1554  
 \bibitem[Ercolano et al.(2008)]{ercolano08}Ercolano, B., Drake, J.~J., Raymond, J.~C., Clarke, C.~C. 2008, ApJ, 688, 398
 \bibitem[Facchini et al.(2017)]{facchini17}Facchini, S., Birnstiel, T., Bruderer, S., van Dishoeck, E.~F. 2017, A\&A, 605, id. A16
 \bibitem[Frank et al.(2014)]{frank14}Frank, A., Ray, T.~P., Cabrit, S., et al. %Hartigan, P., Arce, H.~G., Bacciotti, F.,  Bally, J., Benisty, M., Eisl\"offel, J., G\"udel, M., Lebedev, S., Nisini, B., Raga, A. 2014,
          2014, in  Protostars and Planets VI, eds. H.~Beuther, et al. %R.~S. Klessen, C.~P. Dullemond, T. Henning, 
          (Tucson: University of Arizona Press), 451
 \bibitem[Gaia Collaboration(2016)]{gaia2016}Gaia Collaboration, T. Prusti, J.H.J. de Bruijne, A.,
          et al. 2016, A\&A 595, A1
 \bibitem[Gaia Collaboration(2018)]{gaia2018}Gaia Collaboration, Brown, A.~G.~A., Vallenari, A., et al.  2018, A\&A, in press,
          arXiv 1804.09365v2
 \bibitem[Gorti \& Hollenbach(2009)]{gorti09}Gorti, U.,  Hollenbach, D. 2009, ApJ, 690, 1539  
 \bibitem[Guilloteau et al.(2011)]{guilloteau11}Guilloteau, S., Dutrey, A., Pi\'etu, V., Boehler, Y.
          2011, A\&A, 529, id. A105
 \bibitem[Herczeg et al.(2007)]{herczeg07}Herczeg, G.~J., Najita, J.~R., Hillenbrand, L.~A., Pascucci, I.  2007, ApJ. 607, 509
 \bibitem[Kitamura et al.(1996a)]{kitamura96a}Kitamura, Y., Kawabe, R.,  Saito, M. 1996a, ApJ, 457, 277
 \bibitem[Kitamura et al.(1996b)]{kitamura96b}Kitamura, Y., Kawabe, R.,  Saito, M. 1996b, ApJ, 465, L137
 \bibitem[Lavalley et al.(1997)]{lavalley97}Lavalley, C., Cabrit, S., Dougados, C., Ferruit, P.,
         Bacon, R. 1997, A\&A, 327, 671  
 \bibitem[Podio et al.(2013)]{podio13}Podio, L., Kamp, I., Codella, C., et al. 2013,
          % Cabrit, S., Nisini, B., Dougados, C., Sandell, G., Williams, J.~P., Testi, L., Thi, W.-F., Woitke, P., Meijerink, R., Spaans, M., Aresu, G., M\'enard, F., Pinte, C. 2013,
         ApJ, 766, id. L5 
 \bibitem[Pyo et al.(2003)]{pyo03}Pyo, T-S., Kobayashi, N., Hayashi, M., et al. 2003, ApJ, 590, 340
 \bibitem[Schneider et al.(2013)]{schneider13}Schneider, P.~C., Eisl\"offel, J., G\"udel, M., et al. %G\"unther, H.~M., Herczeg, G., Robrade, J., Schmitt, J.~H.~M.~M. 
           2013, A\&A, 557, A110 
 \bibitem[Schuster et al.(1993)]{schuster93}Schuster, K.~F., Harris, A.~I., Anderson, N., Russell, A.~P.~G. 1993, ApJ, 412, L67 
 \bibitem[Takami et al.(2004)]{takami04}Takami, M.,  Chrysostomou, A., Ray, T.~P., et al. %Davis, C., Dent, W.~R.~F., Bailey, J., Tamura, M., Terada, H. 
          2004, A\&A, 416, 213
 \bibitem[Testi et al.(2002)]{testi02}Testi, L., Bacciotti, F., Sargent, A.~I., Ray, T.~P.,
         Eisl\"offel, J. 2002, A\&A, 349, L31   
 \bibitem[White et al.(2014)]{white14}White, M.~C., McGregor, P.~J., Bicknell, G.~V., Salmeron, R., Beck, T.~L. 2014,
          MNRAS, 441, 1681
\bibitem[White \& Hillenbrand(2004)]{white04}White, R.~J., Hillenbrand, L.~A. 2004,
          ApJ, 616, 998
\bibitem[Yen et al.(2014)]{yen14}Yen, H.-W., Takakuwa, S.,  Ohashi, et al. 2014, ApJ, 793, 1 

\end{thebibliography}

\begin{appendix} %First appendix

\section{First-moment maps for $^{12}$CO}

\begin{figure}[h!]
\hskip -0.1truecm\includegraphics[width=8.93cm, trim={0 0 0 10}, clip]{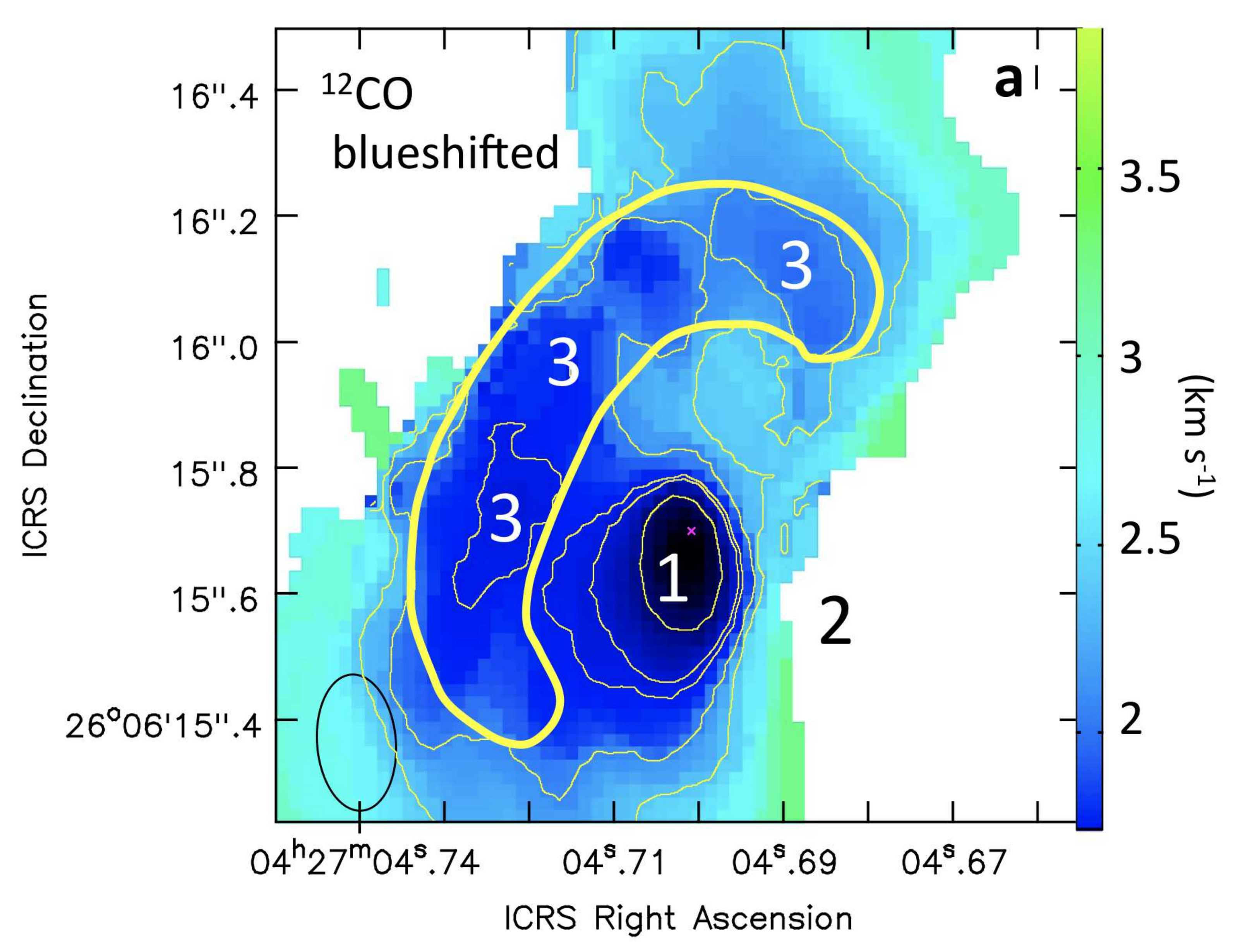}
\hbox{\hskip -0.03truecm\includegraphics[width=9.1cm, trim={0 0 0 10}, clip]{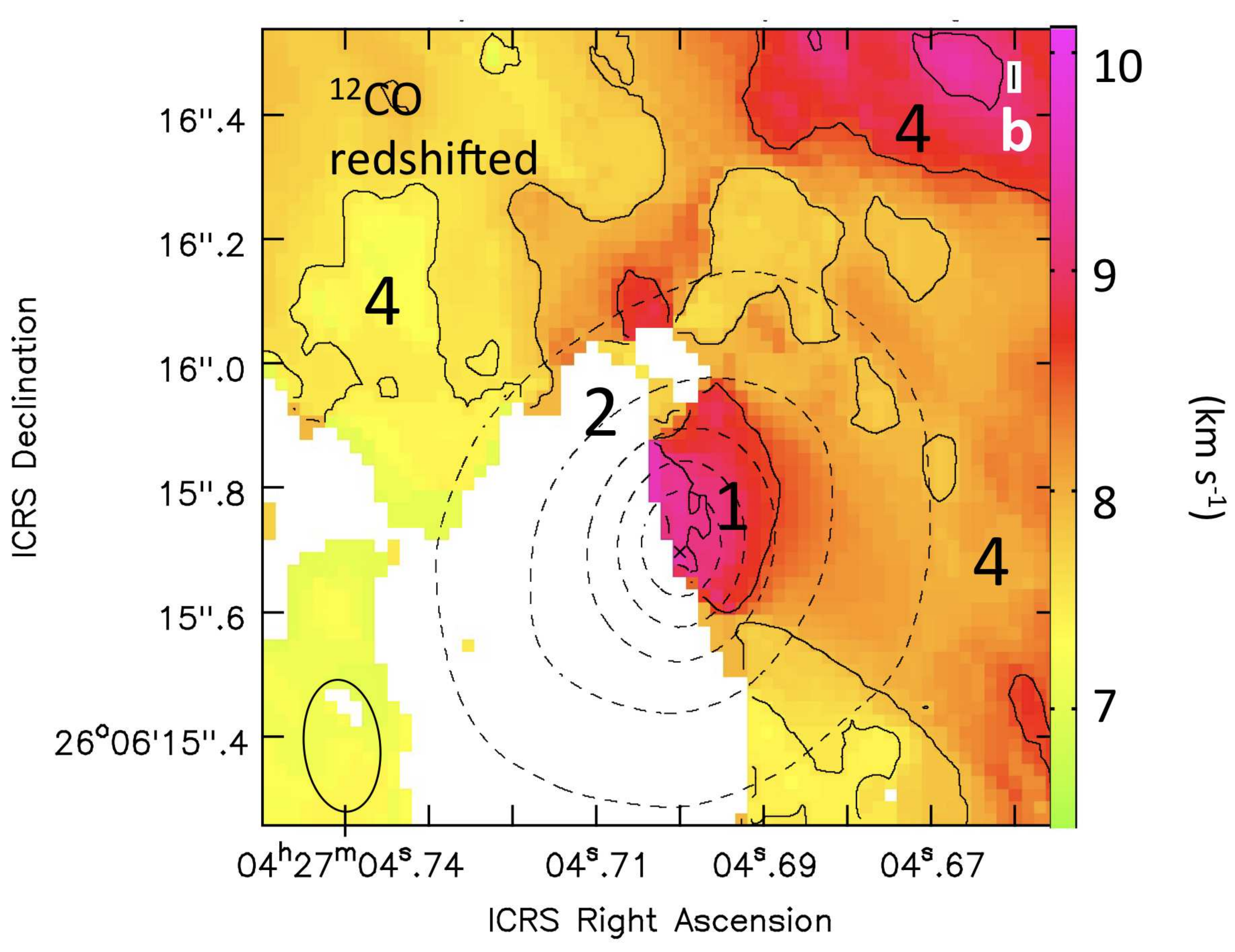}}
\caption{Zoom-in on the area of the continuum disk in Fig.~\ref{fig:12CO}a,b for $^{12}$CO. Labels 1--4 refer to
      the features defined in Sect.~\ref{results}.
      {\bf (a)} \textit{Top}: Blueshifted, 
      $v_{\rm lsr} = $ -1.2 -- +3.6~km~s$^{-1}$ (color; black below 1.6~km~s$^{-1}$).  
      Thin yellow contours are plotted for a few arbitrary velocity levels for visualization.
      The thick yellow outline marks the approximate area of the high-velocity wind. {\bf (b)} \textit{Bottom}:  Redshifted, $v_{\rm lsr} = 7.0-10.1$~km~s$^{-1}$. Thin black contours are overplotted 
      for a few arbitrary velocity levels.
      Beam size: $0\farcs 21$~$\times$~$0\farcs 12$. }\label{zoom}
\end{figure}

We show here more details related to Fig.~\ref{fig:12CO}a,b. Figure~\ref{zoom} zooms in on the innermost 
parts, about coincident with the continuum-emitting dust disk (dashed contours in Fig.~\ref{zoom}b). We label here features
[1] to [4] as defined in Sect.~\ref{results}, and in particular outline the blueshifted arc-like 
structure [3] with a yellow line in Fig.~\ref{zoom}a.

\begin{figure}
\hskip -0truecm\includegraphics[width=9.0cm, trim={0 0 0 0}, clip]{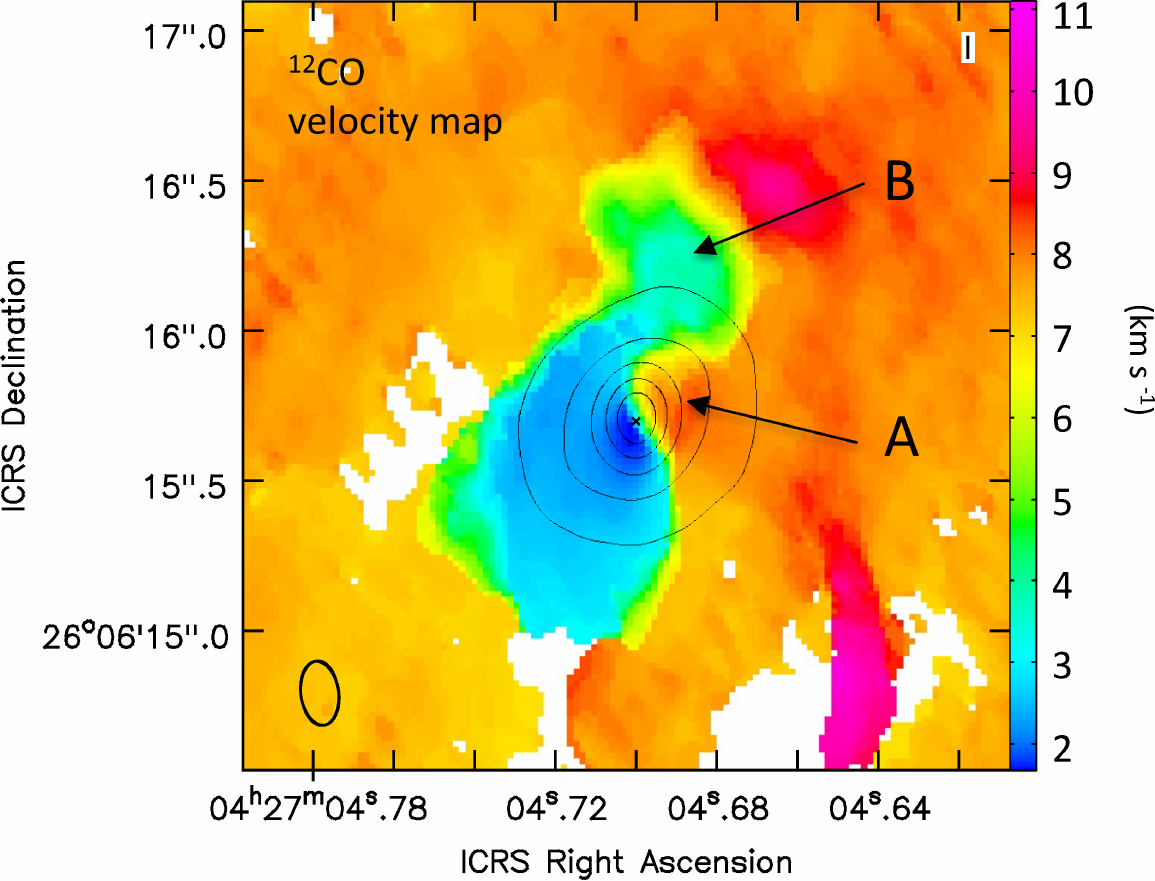}
\caption{First-moment map for $^{12}$CO ($v_{\rm lsr}$=1.5--11~km~s$^{-1}$). 
      Beam size: $0\farcs 21$~$\times$~$0\farcs 12$ (ellipse at lower left). 
      Contours show continuum emission (4.3, 15.1, 26.5, 37.8, 49.2~mJy/beam, peak 60.6~mJy/beam).
      The cross symbol indicates the position of the star. Arrows point to
      problematic areas for this map (see text).
}
\label{moment12CO}
\end{figure}

In Fig.~\ref{moment12CO} we show the combined full first moment map for $^{12}$CO in a similar fashion 
as done in Fig.~\ref{fig:firstmoment}a,b for $^{13}$CO and C$^{18}$O. While this figure shows the general arrangement of 
red- and blueshifted kinematic features,  we caution that independent blue- and redshifted features are superimposed along some
lines of sight. For example, arrow ``A'' points to the inner Keplerian disk from which we measure, expectedly, redshifted emission.
But Fig.~\ref{fig:12CO}a and Fig.~\ref{zoom}a show that there is also a blueshifted component along the same lines of sight,
probably part of the wind. Arrow ``B'' points to a region that emits strongly blueshifted in Fig.~\ref{fig:12CO}a and Fig.~\ref{zoom}a
but is displayed here at a systemic velocity close to rest (green color), as there is also strong redshifted emission in the 
same region. 

\begin{figure}[h!]
\hskip -0.3truecm\includegraphics[width=9.4cm, trim={0 0 0 10}, clip]{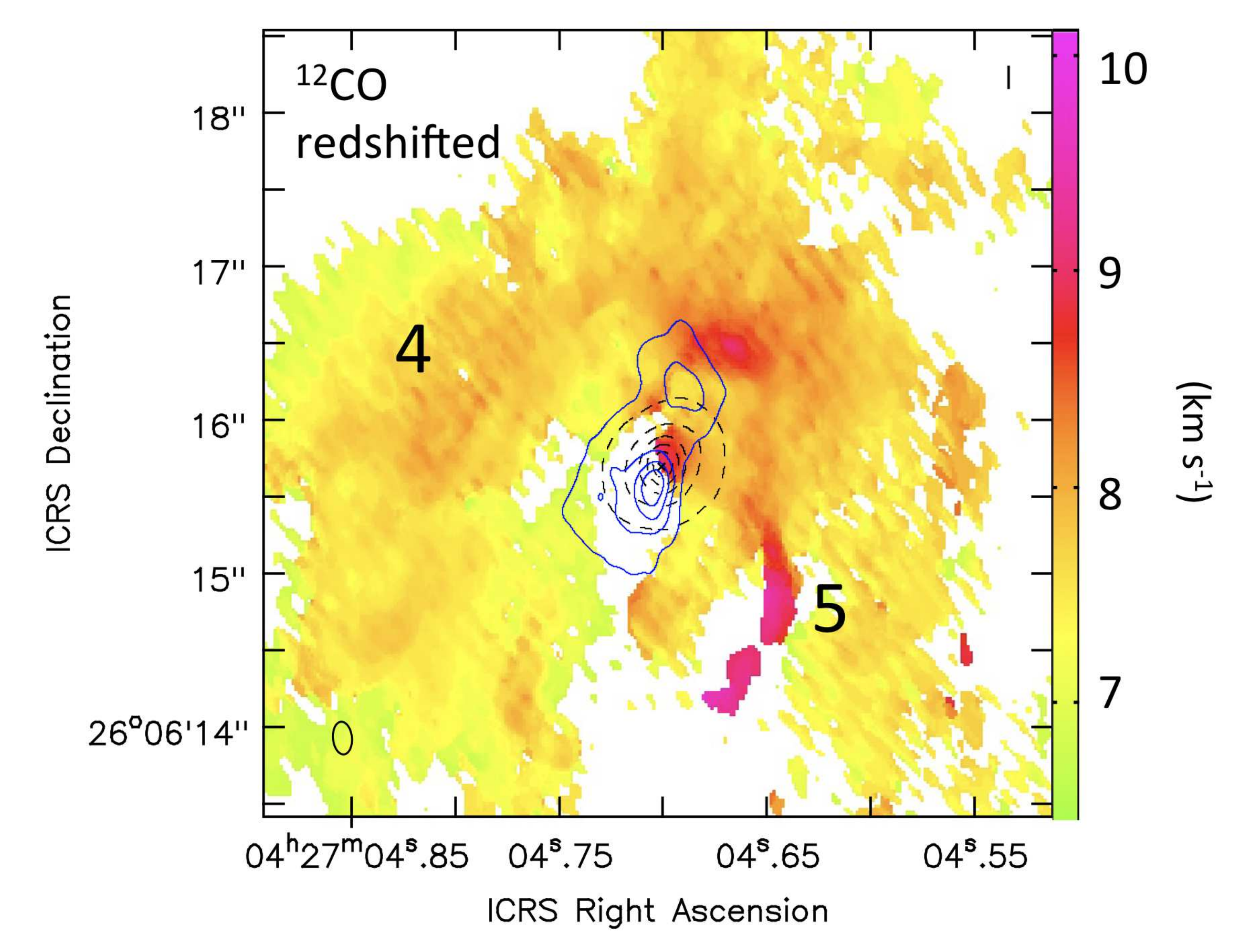}
\caption{Larger field around DG Tau in redshifted channels similar to Fig.~\ref{fig:12CO}b.  We note the arc-like structure [5] at high velocity.}
\label{redlarge}
\end{figure}

In Fig.~\ref{redlarge} we  show the larger field around Fig.~\ref{fig:12CO}b, that is,
the redshifted channels. In particular, the figure shows the arc-like high-velocity structure 
in the SW [5] that is briefly described in Sect.~\ref{discussion}. There are no significant blueshifted sources 
outside the field of Fig.~\ref{fig:12CO}a.

\section{C$^{18}$O and $^{13}$CO intensity maps}\label{absorption}

For completeness, we provide here some complementary 
observational information related to the intensity maps for C$^{18}$O and $^{13}$CO; those of $^{12}$CO
were shown in the main paper (Fig.~\ref{fig:12CO}c,d).

\begin{figure}[h!]
\vbox{
\includegraphics[width=8.9cm,trim={383 220 40 100}, clip]{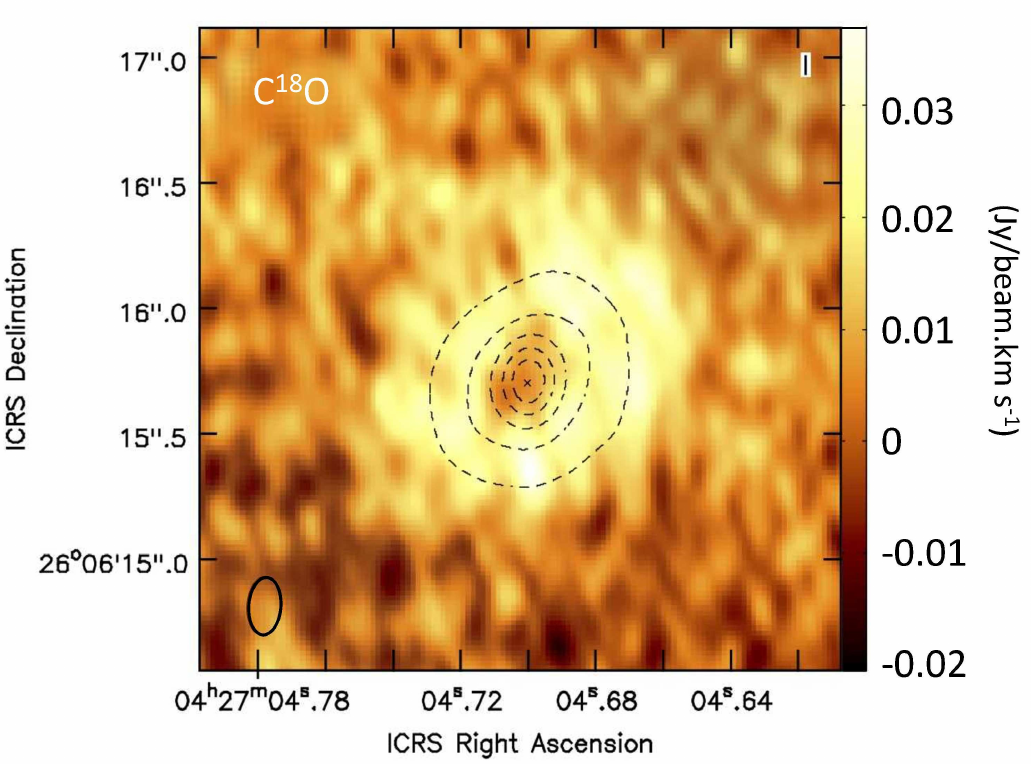}
\includegraphics[width=8.9cm,trim={73 220 350 100},clip]{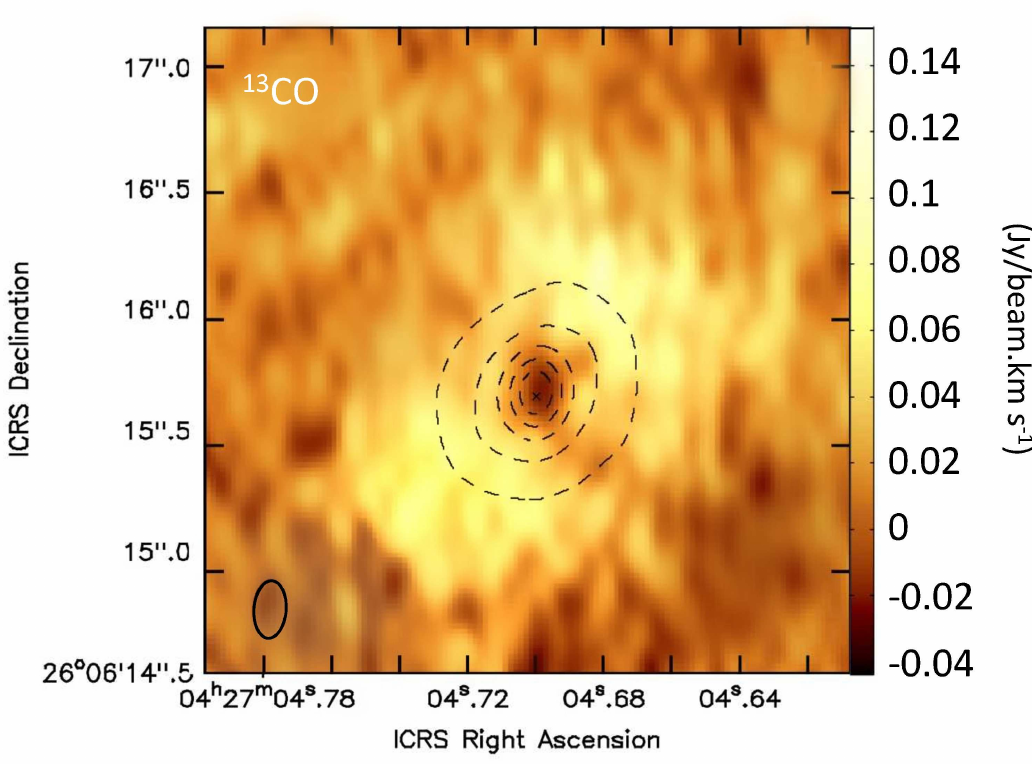}
}
\caption{Intensity maps for (top) C$^{18}$O (top) and $^{13}$CO (bottom) integrated over the entire  velocity
         range of the line. We note  the weak emission from the central regions and the diagonal (NE-SW) absorption band
         in the $^{13}$CO map. Dashed contours represent the same as those in Fig.~\ref{fig:12CO}a,b.
}\vskip -0truecm
\label{absmaps}
\end{figure}

The C$^{18}$O intensity map integrated over all relevant
velocities (Fig.~\ref{absmaps}-top) shows an ellipsoidal ring of emission between radii of  
$\approx 0\farcs 2$ and $0\farcs 6$ along the major axis. The emission inside $0\farcs 2$ is 
faint.  There is no indication for a velocity-specific absorption feature (see  
Fig.~\ref{fig:firstmoment}a). Apart from a real ``hole'' in the disk with depleted gas and therefore CO deficiency, 
various opacity effects could be acting. First, the dust could be optically thick in the inner disk, suppressing line emission
(e.g., \citealt{cleeves16} for IM Lup). For a derivation, one needs observations across a wide range of frequencies, which are
presently not available in the same spatially resolved quality as used in our paper. \citet{guilloteau11} modeled the radial
dependence of the dust opacity of DG Tau using IRAM Plateau de Bure Interferometer data with a resolution down to 
$\sim 0\farcs 4$ at 1.3--3~mm  wavelengths; they found that the disk is optically thick within 40--50~au, compatible
with the holes in our observations that reach out to at least $\sim$25~au. Numerical disk formation models indicate 
optically thick dust disks within about 10 au (Fig.~5 in \citealt{dunham14}).

 On the other hand, the gas could be partially optically thick and absorb some dust continuum emission. 
\citet{boehler17} argued, for a similar observation of HD~142527, that the line 
emission is underestimated where the dust is bright and the dust emission is overestimated,
because at the line frequency molecular gas can partially absorb dust emission. This leads to suppressed 
emission from the central region after subtraction of a constant background
evaluated outside the line region in the spectrum. Which model applies to DG Tau needs more detailed multi-wavelength 
observations and comprehensive thermo-chemical disk opacity modeling.

The $^{13}$CO intensity map (Fig.~\ref{absmaps}-bottom) is similar, showing bright emission mostly in the  
$0\farcs 2$ and $0\farcs 6$ radius range.
In addition however, as already noticed in the first-moment map (Fig.~\ref{fig:firstmoment}b), a diagonal
band from NE to SW suppresses emission along the projected minor axis of the disk, mostly coinciding with an
area for which a Keplerian disk shows low radial velocities. This symmetric feature is very likely related to the radial velocity relative to the position of the observer and seems to reach across the disk diameter.
No bright high-velocity emission is evident in the map (very different from $^{12}$CO in Fig.~\ref{fig:12CO}c,d), 
indicating that like for C$^{18}$O, the inner-disk emission is relatively faint.
We conclude that the $^{13}$CO emission from most likely the entire disk surface is absorbed by a foreground layer of
cool gas at specific, small  velocities relative to $v_{\rm sys}$.

\section{Flow geometry}\label{sect:geometry}

\begin{figure}[h!]
\hskip -1.5truecm\includegraphics[width=10.2cm, trim={0 0 0 70}, clip]{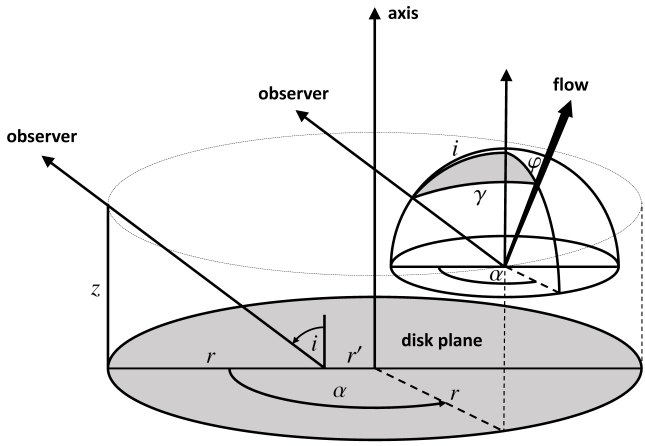}
\vskip -0.5truecm
\caption{Geometry of the flow relative to the disk and the line of sight. The protostellar disk
         midplane is shaded in gray, out to a radius $r$. The line of sight is from the
         upper left; it is inclined by an angle $i$ to the disk axis. On the left side, 
         a source at height $z$ above the disk is projected inward to a radius of 
         $r^{\prime}$, which itself will be compressed to $r^{\prime}\cos i$ due to 
         disk inclination $i$ (see Eq.~\ref{projection}). The half-sphere on the right shows, 
         for a point at projected 
         disk radius $r$ at a height $z$ and azimuth angle $\alpha$, the relative geometry
         between the vertical (parallel to the disk axis), the line of sight to the
         observer (inclination $i$), and the flow direction (angle $\varphi$ away from the
         disk axis; see Eq.~\ref{proj}). The spherical triangle shaded in gray is used to
         calculate the angle between the flow direction and the line of sight.}
\label{geometry}
\end{figure}
Figure~\ref{geometry} shows the adopted geometry of the star-disk-jet system of DG Tau together
with the geometry of the flow relative to the axis and the line of sight. The left half of the
diagram explains the geometry of projection of the blueshifted flow emission onto the plane of
the disk (Eq.~\ref{projection}). We assume a flow emitting at a height $z$ above the 
disk midplane. The half-sphere on the right describes the geometry of the flow and defines angles
between the disk rotation axis, the line of sight, and the flow direction
at the position of the sphere center, at a height $z$ above the disk plane at azimuth $\alpha$. Here, we assume 
an axisymmetric flow with an outward inclination angle of $\varphi$ from the local vertical.

\end{appendix}

\end{document}